\documentclass[sigconf,natbib=false]{acmart}

\usepackage{comgipp}
\usepackage[
	backend=biber,
	style=numeric,
	firstinits=true,
	url=false,
	isbn=false,
	]{biblatex}
\newcommand{\theReferenceText}{\fullcite{Schubotz2018b}}
\usepackage{graphicx}\usepackage{subfig}
\copyrightyear{2018}
\acmYear{2018}
\setcopyright{usgovmixed}
\acmConference[JCDL '18]{The 18th ACM/IEEE Joint Conference on Digital Libraries}{June 3--7, 2018}{Fort Worth, TX, USA}
\acmBooktitle{JCDL '18: The 18th ACM/IEEE Joint Conference on Digital Libraries, June 3--7, 2018, Fort Worth, TX, USA}
\acmPrice{15.00}
\acmDOI{XXX}
\acmISBN{978-1-4503-5178-2/18/06}

\begin{document}

\title{Introducing MathQA -\\
  A Math-Aware Question Answering System}
\author{Moritz Schubotz, Philipp Scharpf,\\
  Kaushal Dudhat, Yash Nagar, Felix Hamborg, Bela Gipp}
\affiliation{
    \institution{
    Information Science Group, University of Konstanz, Germany (first.last@uni-konstanz.de)}}
\renewcommand{\shortauthors}{Schubotz and Scharpf}

\begin{abstract}
We present an open source math-aware Question Answering System based on Ask Platypus.
Our system returns as a single mathematical formula for a natural language question in English or Hindi.
This formulae originate from the knowledge-base Wikidata.
We translate these formulae to computable data by integrating the calculation engine sympy into our system.
This way, users can enter numeric values for the variables occurring in the formula.
Moreover, the system loads numeric values for constants occurring in the formula from Wikidata.
In a user study, our system outperformed a commercial computational mathematical knowledge engine by 13 \%.
However, the performance of our system heavily depends on the size and quality of the formula data available in Wikidata.
Since only a few items in Wikidata contained formulae when we started the project, we facilitated the import process by suggesting formula edits to Wikidata editors.
With the simple heuristic that the first formula is significant for the article, 80 \% of the suggestions were correct.
\end{abstract}
\maketitle
\thispagestyle{firststyle}

\begin{figure}
\includegraphics[width = \columnwidth]{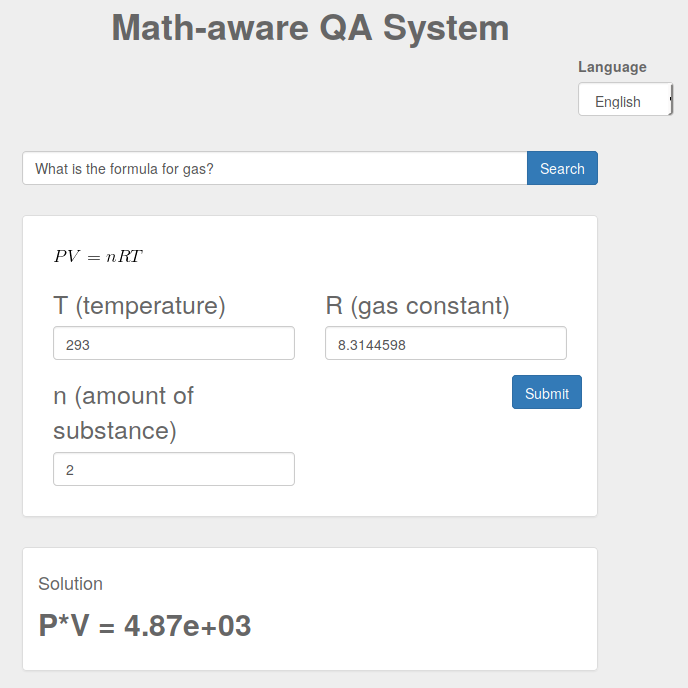}
\caption{Screenshot of MathQA}
\end{figure}

\section{Introduction}

Question Answering (QA) systems are Information Retrieval (IR) systems, allowing the user to pose questions in natural language to provide quick and succinct answers - in contrast to search engines which deliver ranked lists of documents.
In this project, we developed an open source QA system, which is available at \url{https://github.com/ag-gipp/MathQa}.
Our system can answer mathematical questions in the form of natural language, yielding a formula, which is retrieved from Wikidata. Wikidata is a free and open knowledge-base that can be read and edited by humans and machines. It stores common sources of other Wikimedia projects, especially for Wikipedia infoboxes.
In addition, our system enables the user to perform arithmetic operations using the retrieved formula.
We developed three modules:
The \emph{Question Parsing Module} (1) transforms questions into a triple representation and produces a simplified dependency tree.
The \emph{Formula Retrieval Module} (2) then queries the Wikidata knowledge-base for the requested formula and presents the result to the user.
The user can subsequently choose values for the occurring variables and order a calculation that is done by a \emph{Calculation Module} (3).
If available, the system retrieves the identifier names and values from Wikidata, so that the user can understand their meaning (see Figure \ref{fig:QAS_GUI} above).
Moreover, we developed a module which can answer questions in the Hindi language. In contrast to the English, which is exploiting the dependency graph of Stanford NLP, the Hindi module uses regular expressions to parse the questions and before passing the triple representation to the Formula Retrieval Module.
Our QA system builds upon Ask Platypus \cite{AskPlatypus}, an existing QA engine that can answer English questions using Wikidata.
We chose Ask Platypus as the best among other Wikidata-based QA systems and extended its functionality to include mathematical questions.
Finally, we evaluated the system's performance and the quality of results in comparison to a commercial computational mathematical knowledge-engine.
Our system outperforms the reference engine on some definition and geometry type questions, and we conjecture that the validity can be expanded to the whole domain.
Before building the QA system, we performed a seeding of all currently available mathematical formulae (labeled by math-tags) from Wikipedia into Wikidata.
Each section of this paper is divided into two parts: the first part describes the process of seeding the Wikidata knowledge-base with mathematical formulae from Wikipedia as a separate project, laying the foundation for the second part, its application in the QA system.

\paragraph{Vision}

The mathematical QA system is a first motivating application that exploits the mathematical knowledge seeded into Wikidata. It is a first step towards our long-term goal of building a \textit{collaborative, semi-formal, language independent math(s) encyclopedia} hosted by Wikimedia at \textit{math.wikipedia.org} \cite{corneli2017math}. Using the popular Wikipedia framework as frontend will help popularize the project and motivate many experts from the mathematical sciences to contribute. We envisage a future centralized, machine-readable repository for mathematical world-knowledge that can be utilized to enable cross-article queries, e.g., to automate proofs of mathematical theorems. A crucial foundation for a path towards this long-term goal is having a large amount of mathematical data in Wikidata. This paper is a starting point for the development of effective methods to automatically seed Wikidata with mathematical formulae from Wikipedia or STEM documents.

\paragraph{Problem Setting.}

Wikipedia consists of many pages related to mathematics. However, promptly grasping the essence of an article can be a difficult task as many pages contain a lot of information. Using Wikipedia means reading articles, and there currently is no way to automatically gather information scattered across multiple articles \cite{DBLP:journals/ws/KrotzschVVHS07}. To overcome this problem, Wikidata can be used as a source. Wikidata provides machine-readable content that can automatically be interpreted by a computer and queried to access specific information. Thus, there is a huge potential in adding formulae related to all mathematical topics as items to Wikidata, enabling direct access to the defining formula of a requested mathematical concept. The first goal of this project was to enrich Wikidata with mathematical knowledge it currently lacks. Adding this information into Wikidata will not only increase the content of these items but also make them more meaningful and useful. Furthermore, these formulae will be machine-interpretable and can be used in many applications in the STEM disciplines. Most importantly, we are then able to develop the mathematical QA system which can directly answer mathematical questions provided by the user, using the mathematical formulae and relations available on Wikidata. As a result, instead of retrieving a whole Wikipedia page which is full of text, users will directly get the desired piece of information, the formula they are looking for.

\paragraph{Research Objectives.}

Motivated by the lack of mathematical knowledge in Wikidata, the following research objective was defined:

\begin{center}
\textbf{Identify and extract defining formulae from all the available mathematical articles on Wikipedia to seed them into the Wikidata knowledge-base.}
\end{center}

To achieve this objective, the following tasks were performed: 1.) Identification of mathematical articles from the Wikipedia data dump. 2.) Manual analysis to determine the defining formula of an individual article. 3.) Seeding of the retrieved formulae into Wikidata using the \textit{Primary sources tool} \cite{PrimarySources}. 5.) Evaluation of the overall correctness and accuracy of the data migration by precision, recall, and f-measure.

Subsequently, we capitalized on the formulae seeded into Wikidata to

\begin{center}
\textbf{Build a math-aware QA system, processing a mathematical natural language question to retrieve a formula from Wikidata and allow a calculation based on input values for the occurring variables provided by the user.}
\end{center}

We performed the following subtasks: 1.) Development of a \textit{Question Parsing Module} that determines a triple representation of the user's input. 2.) Development of a \textit{Formula Retrieval Module} to query Wikidata using pywikibot \cite{Pywikibot}. 3.) Development of a \textit{Calculation Module} that performs a calculation based on the retrieved formula for the question and input values for the variables provided by the user. 4.) Evaluation of the overall performance and comparison to a commercial computational mathematical knowledge-engine. 5.) Development of regular expressions to maximize the number of answerable questions provided by the user in the Hindi language.

\paragraph{Section Outline.}

This paper is organized as follows: Section \textit{Background} contains details about the Wikimedia sister projects Wikipedia and Wikidata and the concept of QA systems. Subsection \textit{Implementation} describes our approach of transferring formulae from Wikipedia to Wikidata and the structure of the QA system which uses the seed. In subsection \textit{Evaluation} we describe the construction of a random sample to assess the quality of the data transfer by precision, recall and f-measure. Subsequently, we evaluate the performance of the QA system and discuss its limitations. Finally, we conclude with a summary and suggested improvements for future work.

\section{Background} %
\pagestyle{standardpagestyle}
\subsection{Wikipedia and Wikidata}

Started in 2001, mainly as a text-based resource, Wikipedia\footnote{\url{http://www.wikipedia.org}} is the world's largest online encyclopedia which allows its users to edit articles and add new information into it \cite{wiki:Wikipedia}. Wikipedia has collected an rapidly increasing amount of information, including numbers, coordinates, dates and other types of relationships among different domains of knowledge. Denny Vrandecic, ontologist at Google, claims that \textit{It has become a resource of enormous value, with potential applications across all areas of science, technology and culture} \cite{DBLP:journals/cacm/VrandecicK14}.

Wikipedia is open and welcomes everyone who wants to make a positive contribution. Ward Cunningham, the inventor of Wiki, describes Wikipedia as \textit{The simplest online database that could possibly work} \cite{leuf2001wiki}.

The following are some characteristics of Wikipedia which enable it to manage its data on a global scale:

\begin{itemize}
\item Open and instantaneous editing: Wikipedia allows its users to extend and edit the available information even without creating an account. All changes are instantaneously released online.
\item Record of editing history: Wikipedia keeps a record of all the changes made to a page. The page history can be viewed by everyone. Each time a page is edited, the new version is released, and the old version is saved in the page history.
\item Linked pages: all Wikipedia pages are linked to other Wikipedia pages that are related to each other which results in a web of interlinked pages.
\item Multilingual: Wikipedia exists in many languages. Every article on Wikipedia consists of a list of languages it is available in.
\item Content standard: the information contributed to Wiki\-pedia must be encyclopedic, neutral and verifiable.
\item Community control: Wikipedia is always supported by a team of dedicated volunteers who take the responsibility of developing content, policies, and practices.
\item Continuous evolution: Wikipedia is always in a state of continuous growth. Information is continuously being added and updated and new features are added into Wikipedia to make it more useful.
\item Totally free: all Wikipedia content is free to use, anyone is free to contribute, and the content is released under a free license which means anyone may reuse it elsewhere. Wikipedia is a non-commercial project, and it has no advertisements.
\end{itemize}

Although Wikipedia comprises a huge amount of data, it does not provide direct access to specific facts, as it is still unstructured, which is unfortunate for anyone who wants to retrieve information systematically. To remedy this shortcoming, the Wikimedia Foundation launched Wikidata\footnote{\url{http://www.wikidata.org}} in October 2012. Wikidata is a free and open knowledge-base that can be read and edited by humans and machines. It acts as a common source of data which can be used by Wikimedia projects such as Wikipedia, Wikivoyage, Wikisource, and others. As Wikidata is one of the most recent projects of the Wikimedia Foundation, it is still in an early phase of development. Therefore, it encourages its users and organizations to donate data so that it can grow and distribute open source multilingual and educational content free of charge. Wikidata's data model basically consists of item and statement. Each item represents an entity, such as a person's name, a scientific term, a mathematical theorem, etc. and is identified by a unique number prefixed by the letter Q. For instance, the item number (QID) for the topic \textit{Computer science} is \textit{Q21198}. Additionally, items may have labels, description, and aliases in multiple languages. Information is added to items by creating statements and stored in the form of key-value pairs with each statement consisting of a property as a key and a value associated with that property.

\begin{figure}
\includegraphics[width = \columnwidth]{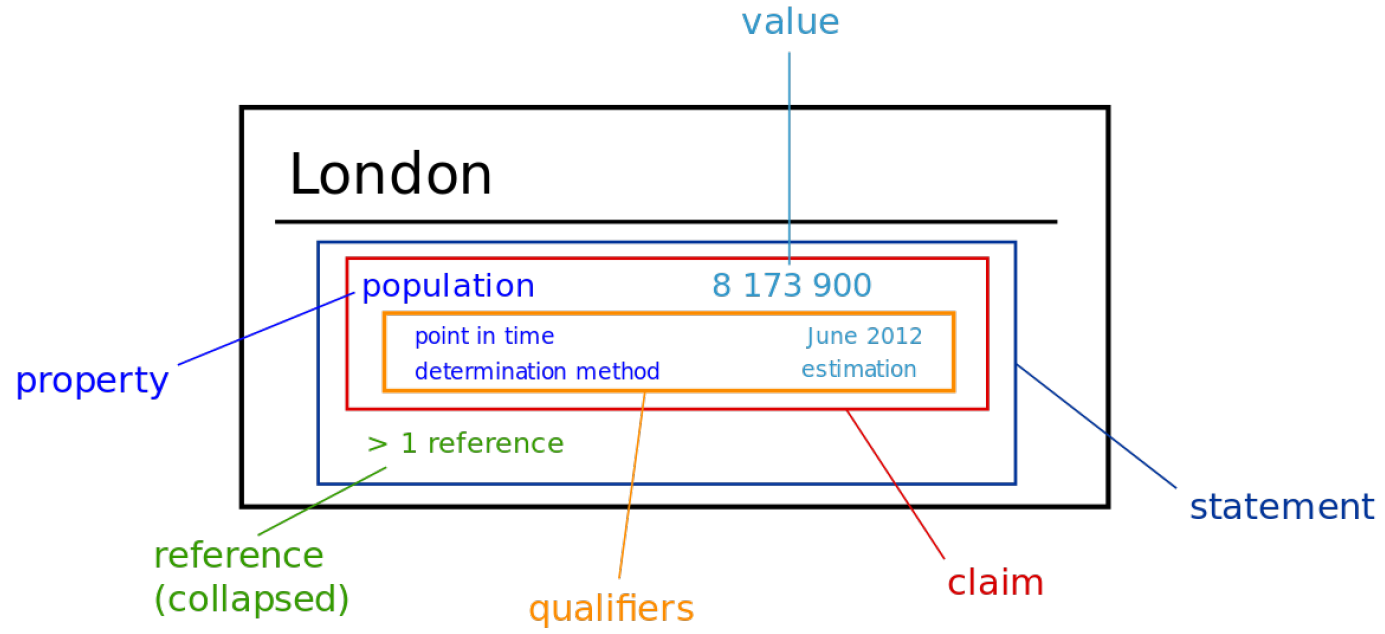}
\caption{Wikidata statement terminology illustrated by an example [11].}
\label{fig:Wikidata_statement_terminology}
\end{figure}

Figure \ref{fig:Wikidata_statement_terminology} illustrates the data model used in Wikidata.
In this example, \textit{London} is the main item with a statement which consists of one claim and one reference for the claim. The claim itself contains a main property-value pair which represents the main fact, which is \textit{population} and the corresponding value in this case. Some optional qualifiers can also be added into the claim to append additional information related to the main property. In this example, the time at which the population was recorded and the \textit{determination method} are the qualifiers with their corresponding values \textit{June 2012} and \textit{estimation}, respectively. Wikidata does not claim to provide users with true facts because many facts are disputed or uncertain. Therefore, Wikidata allows such conflicting facts to coexist so that different opinions can be expressed properly. The content of Wikidata is in public domain under Creative Commons CC0 license which allows every user to use, extend and edit the stored information. The Wikidata requirements \cite{WDRequirements} state that the data uploaded to Wikidata must not overwhelm the community. Research in understanding how systems for collaborative knowledge creation are impacted by events like data migration is still in its early stages \cite{DBLP:journals/ce/MoskaliukKC12}, in particular for structured knowledge \cite{DBLP:journals/bioinformatics/HorridgeTNVNM14}. Most of the research is focused on Wikipedia \cite{flock2015towards}, which is understandable considering the availability of its data sets, in particular the whole edit history \cite{DBLP:journals/expert/SchindlerV11} and the availability of tools for working with Wikipedia \cite{DBLP:journals/ai/MilneW13}.

\subsection{Question Answering Systems}

There has been a significant rise in the usage of open-domain QA systems since the establishment of the question answering track in the Text Retrieval Conferences, beginning with TREC-8 in 1999 \cite{DBLP:journals/nle/Voorhees01}. However, in 1960, Simmons published a survey article named \textit{Answering English Questions by Computer} and a paper which describes more than fifteen English language QA systems that were implemented in the previous five years \cite{DBLP:journals/cacm/Simmons65} \cite{DBLP:journals/tois/KwokEW01}. QA systems which rely on database approaches are briefly described in \cite{DBLP:journals/nle/HirschmanG01}. The quality of a query for an IR system has a direct impact on the success of the search outcome. In fact, one of the most important, but frustrating tasks in IR is query formulation \cite{DBLP:conf/cikm/FrenchPGP01}. Relevance feedback strategies, selecting important terms from documents that have been identified by the user as \textit{relevant}, are frequently used \cite{salton1997improving}. Various tools are available which answer queries from diverse fields. The START \cite{START} Natural Language Question Answering System aims to supply users with \textit{just right information} instead of merely providing a list of hits. The NSIR \cite{DBLP:conf/lrec/RadevQWF02} Web-based question answering system implemented a method called \textit{Probabilistic Phrase Reranking} (PPR) where a potential answer that is spatially close to the question words gets a higher score. Some QA systems are domain-specific and focus on a specialized area such as HONQA (Health On the Net Foundation) \cite{HONQA} for quality health care questions. LAMP \cite{zhang2003web} is a QA system which uses snippets from the results returned by a search engine like Google. This property is called \textit{snippet tolerant} and is useful because it can be time-consuming to analyze and download the original web documents. There are some tools available online that perform arithmetic operations and also answer general and mathematical questions, commercial computational mathematical knowledge-engines. A search engine which harvests the web for content representation of mathematical formulae and marks them with \textit{substitution tree indexing} was implemented by Kohlhase and Sucan \cite{DBLP:conf/aisc/KohlhaseS06}. There has been a lot of research related to the extraction of mathematical formulae and discovering relations between a formula and its surrounding text \cite{DBLP:conf/mkm/PagelS14} \cite{DBLP:conf/amia/LeeCZSSEY06} \cite{yokoi2011contextual} \cite{quoc2010mining}. Our QA system for mathematical problems using Wikidata will be the first of its kind. However, there are some already existing QA systems which use Wikidata to answer questions from other domains such as Ask Wikidata! \cite{AskWikidata} and Ask Platypus \cite{AskPlatypus} - similar tools that can be used to find any general information from Wikidata by parsing the natural language query entered by the user.

\section{Implementation}

\begin{figure}
\centering
\includegraphics[scale=0.75]{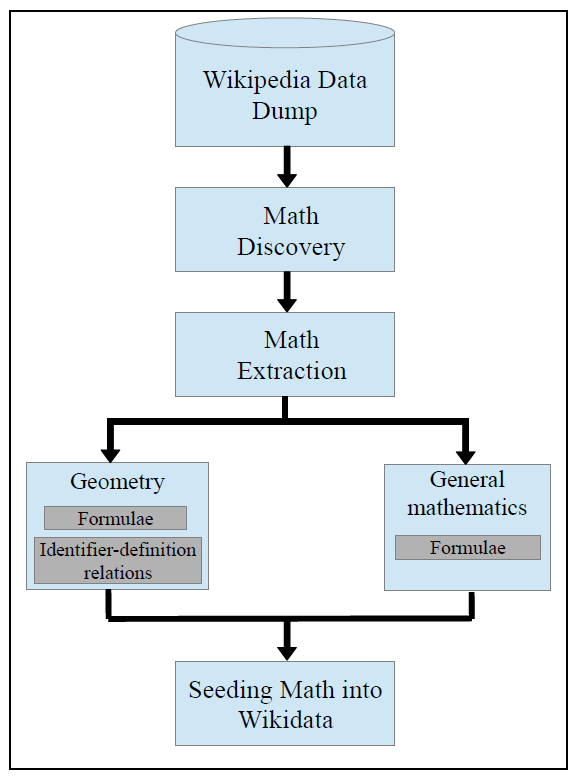}
\caption{Workflow of our extraction and loading process.}
\label{fig:extraction_workflow}
\end{figure}

This chapter describes the implementation details of the formula seeding and QA system.
Figure \ref{fig:extraction_workflow} shows the workflow of the data transfer from Wikipedia to Wikidata.
The first task was to download the Wikipedia data dump and identify the articles that are related to mathematics. Subsequently, we needed to determine and extract the defining formulae from each article. The extraction process was divided into two categories. We distinguished articles related to geometry from the rest of general mathematics. After the extraction, the formulae were added into Wikidata using its Primary sources tool \cite{PrimarySources} that allows users to approve or reject a claim and its reference.

\subsection{Seeding Math into Wikidata}

As mentioned before, Wikipedia consists of millions of documents in various languages. Since its content is barely machine-interpretable, we needed to find a way to distinguish mathematical articles from the rest. To achieve this, the first step was to reduce the number of articles. We only considered English Wikipedia as the primary dataset to profit from two advantages. Firstly, it reduced the total number of articles from over 40 to around 5,5 million. Secondly, English Wikipedia contains the highest number of articles compared to other languages. So, we assume that a mathematical article available in any other language is also available in English Wikipedia, however, vice-versa it might not be true in numerous cases. To recover the mathematical formulae, we needed to distinguish formulae from their surrounding text by identifying the \verb|<math>| tags. Performing the math tag search, we found 32.682 pages containing math formulae After the discovery of mathematical articles in the English Wikipedia dataset, we were confronted with a more significant challenge: How to determine the \textit{defining formula} within a mathematical article. A Wikipedia page contains all the information related to a particular topic. Mathematical pages often contain derivations along with an equation. If we extract all the equations included in math tags, we get a lot of unrelated and irrelevant formulae. To solve this problem, we found a simple yet effective solution; we came up with after manually analyzing a set of random mathematical articles. We observed that in most of the mathematical Wikipedia articles, the first formula was the most important one related to that topic. As the approach gave false results for the Wikipedia articles related to geometry, we divided the math articles into the categories \textit{general mathematics} and \textit{geometry}. We then used different approaches to math extraction for both categories.

\subsubsection{Geometry Questions.}

The main reason behind separating geometry related articles from the rest is that the structure of these articles is different from those of general mathematical articles. A geometric object has various properties such as volume, area, perimeter, etc. Thus, a single topic can have more than one property and multiple defining formulae. The Wikipedia articles of such geometric shapes (cube, circle, ellipse, etc.) may contain the formula of these properties within different subsections of the article. To solve this problem, we first identified all the pages related to geometry by using a list of 16 Wikipedia geometry categories: \textit{Elementary geometry, Theorems in geometry, Polygons, Elementary shapes, Quadrilateral, Area, Volume, Conic sections, Geometric centers, Circles, Curves, Surfaces, Cubes, Platonic solids, Polytopes and Euclidean plane geometry}. We discovered 292 pages belonging to these categories, each containing multiple relevant formulae. We subsequently retrieved the first formula from each of these subsections. However, not all the subsections of the page provided defining formulae related to the topic. For further refinement, we used a simple keywords based filtration of the following property names: \textit{Area, Volume, Circumference, Perimeter, Circumradius, Inradius and Median}, we considered most important for describing 2- and 3-dimensional shapes. These properties have a unique defining formula that can easily be checked for its correctness in the evaluation. We are strictly limited to adding only one defining formula for each property into Wikidata. As a result, we got 65 formulae for the properties mentioned above belonging to 49 Wikipedia articles related to geometry.

\subsubsection{General Formulae.}

As stated previously, we discovered 32.682 pages in English Wikipedia which are related to mathematics. Out of these, 292 were filtered out as a separate category of geometry. For the remaining pages, we chose a different formula retrieval approach. We extracted the first formula from each Wikipedia article, as in most cases this, in fact, yielded the defining formula instead of, e.g., parts of a derivation or proof. After the discovery and extraction of math formulae from Wikipedia, we handed the list to Wikimedia who seeded the formulae to the Primary sources \cite{PrimarySources} where they can now be approved or rejected by Wikidata users.

\subsection{Building the Math QA system}

Having the formulae seeded into Wikidata, we could build our Math-aware QA system. It consists of three modules written in Python that will be described in the following.

The main aim of the \textit{Question Parsing Module} is to transform questions into a tree of triples - producing a simplified and well-structured tree, without losing relevant information about the original question that was provided by the user. This is done by analyzing the grammatical structure of the question, mapping it into a normal form. For our module, we used the simplified dependency tree representation output of the Stanford Parser \cite{StanfordParser}.

Receiving the triple representation from the Question Parsing Module, the \textit{Formula Retrieval Module} is responsible for extracting formulae from Wikidata using Pywikibot \cite{Pywikibot}, a python library, and collection of tools that automate the work on Mediawiki sites. Typically, the triple representation (subject, predicate, object) is incomplete, with either a missing predicate or object. Once the Wikidata item for the subject is available, the module tries to retrieve the value of the predicate. There are two cases for the values of the predicates.

In the first case, if the value of the predicate is \textit{formula}, Pywikibot looks for the value of the Wikidata property named \textit{defining formula} (P2534) and, if available, replaces the triple object with its value. For instance, \textit{What is the formula for Pythagorean theorem?} has the triple representation (Pythagorean theorem, formula, ?). The module maps the subject of the triple to the Wikidata item and returns the value of the \textit{defining formula property} as object.

In the second case, if the value of the predicate is in our list of geometry properties (\textit{volume, area, radius} etc.), Pywikibot looks for the value of the predicate in the \textit{has quality} property (P1552) of the subject and, if available, replaces the triple object with the \textit{defining formula} (P2534) value. For instance, \textit{What is the volume of a sphere?} has the triple representation (sphere, volume, ?). The module maps the subject to the Wikidata item, the predicate to its \textit{has quality} property and returns the value of the \textit{defining formula} property as object.

The \textit{Calculation Module} module is responsible for calculating the result of the formula, with values for the occurring variables provided by the user. If the names and values of the identifiers are available on Wikidata as \textit{has part} (P527) property, they are automatically retrieved and displayed, so that the user can understand their meaning before entering values. Once the formula is received from the Formula Retrieval Module, it is parsed from LaTeX to Sympy form using the process\_sympy parser \cite{latex2sympy} to subsequently have its identifiers extracted for the calculation that is done using the python library Sympy \cite{SymPy}. In addition to the definition and geometry questions, our system also allows a formula as a question input to provide a calculation based on values for the identifiers.

\begin{figure}
\centering
\subfloat[English question]{
\includegraphics[width=\columnwidth,trim=0 3cm 0 1.3cm,clip]{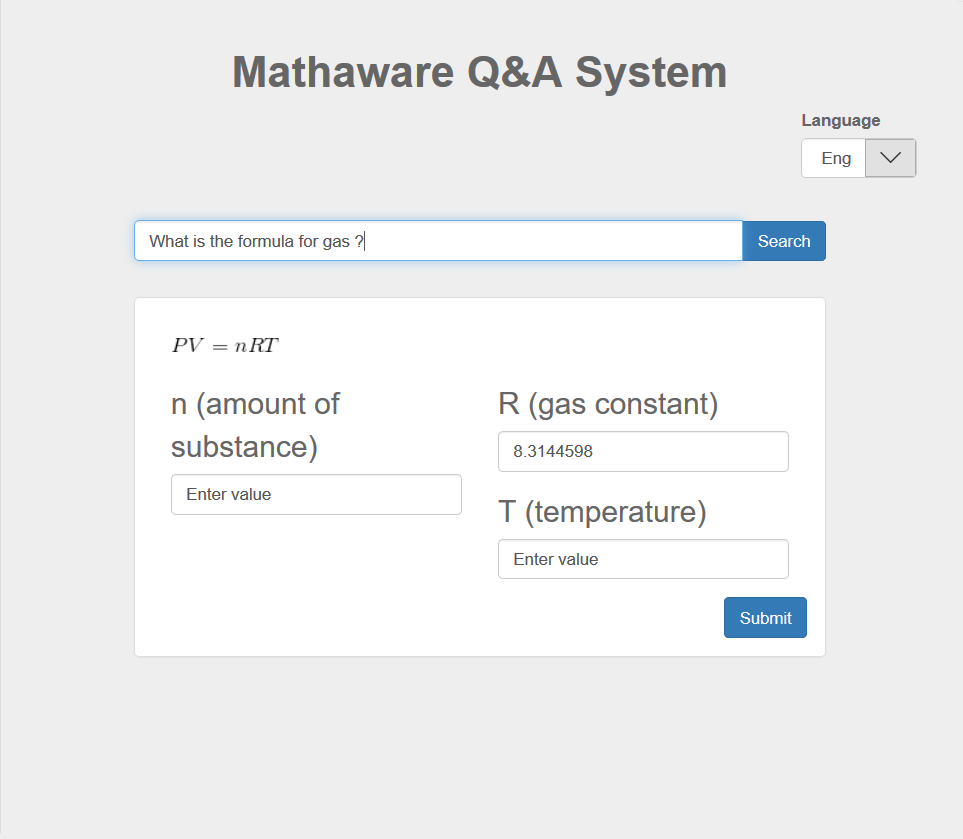}
}
\\
\subfloat[Hindi question]{
\includegraphics[width=\columnwidth,trim=0 0 0 .7cm,clip]{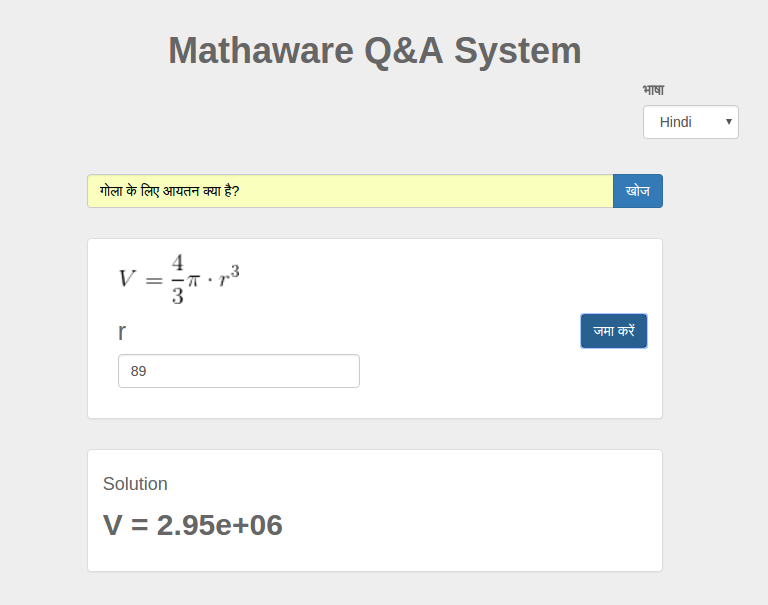}
}\\
\subfloat[Formula question]{
\includegraphics[width=\columnwidth,trim=0 0cm 0 0cm,clip]{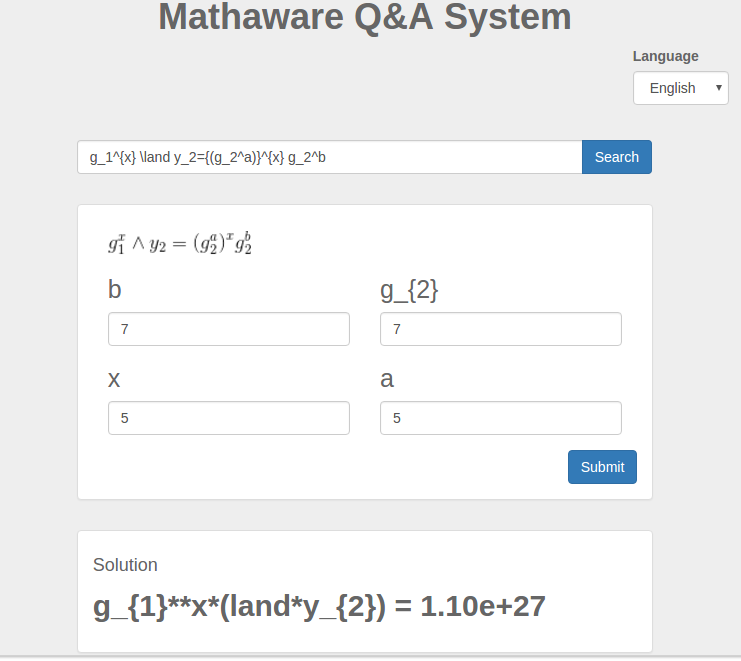}
}
\caption{MathQA GUI for English (a) and Hindi (b) questions, and a direct formula question (c).}
\label{fig:QAS_GUI}
\end{figure}

Figure \ref{fig:QAS_GUI} shows the user web interface (GUI) for English and Hindi questions, as well as a direct formula question.

\section{Evaluation}

Finally, the individual success of the Wikidata seeding and our QA system was evaluated by the standard Information Retrieval measures precision, recall and the combined f-measure.

\subsection{Evaluation of the Wikidata Seeding}

The main goal of the evaluation was to determine how effectively and accurately we were able to retrieve the mathematical formulae from Wikipedia. The evaluation was carried out separately for general mathematics and geometry.

\subsubsection{General Mathematics.}

We evaluated the success of the data transfer by precision and recall, while classifying a result as \textit{relevant} if and only if the retrieved result was estimated to be one and the only general mathematical representation of the Wikipedia article it was extracted from and \textit{non-relevant} if it was not the defining formula or incomplete.
As the formulae were extracted from Wikipedia which is not machine-interpretable, we needed to check \textit{manually} whether a formula is relevant or non-relevant. Since it would have been very exhaustive and time-consuming to check all the formulae extracted from 32.682 Wikipedia pages, we chose a random sample of an arbitrary size of 100, which can be evaluated in a moderate amount of time. We manually examined the Wikipedia articles to find out whether there were defining formulae available for the given mathematical concepts and compared them to the alleged defining formulae of the Wikidata items.
To calculate precision and recall, we classified a result as \textit{relevant} if there was a defining formula of the mathematical concept in its Wikipedia article and \textit{retrieved} if it was the first formula that was subsequently seeded to Primary Sources.
Table \ref{tab:general_evaluation} shows a snapshot of the evaluation of our random sample comprising 100 formulae with their contingencies, whereas Table \ref{tab:Contingency_general} contains the evaluation results for the general formula seeding. The complete list is available at \url{https://github.com/ag-gipp/MathQa}.\\

\begin{table}
\caption{Evaluation of the seeding of general formulae on the basis of a random sample comprising 100 formulae.}
\label{tab:general_evaluation}
\centering
\bgroup
\def\arraystretch{1.8}
\resizebox{\columnwidth}{!}{
\begin{tabular}{|c|c|c|c|c|}
\hline
\textbf{Number} & \textbf{Wikipedia title} & \textbf{Wikidata item} & \textbf{Retrieved result} & \textbf{Contingency} \\
\hline
1 & Holonomy & Q907926 & $\mathrm{Hol}_{x}(\nabla) = \{P_\gamma
\in \mathrm{GL}(E_{x})$ & fp \\
 & & & $|\gamma$ is a loop based at $x\}$ & \\
\hline
2 & Nome & Q7048497 & $f(z)=z^{T}(Mz+q)$ & fn \\
\hline
3 & Jordan's lemma & Q1816932 & $C_{R} = R \cdot e^{i\Theta } | \theta \in [0,\Pi ]$ & fp \\
\hline
... & ... & ... & ... & ... \\
\hline
49 & Matching (graph theory) & Q1065144 & $|A \setminus B| \le 2|B \setminus A|$ & fp \\
\hline
50 & Gaussian function & Q1054475 & $f\left(x\right) = a e^{- { \frac{(x-b)^2 }{ 2 c^2} } }$ & tp \\
\hline
51 & Reisz mean & Q2152569 & not available & fn \\
\hline
... & ... & ... & ... & ... \\
\hline
98 & Plastic number & Q2345603 & $x^3 = x + 1$ & fp \\
\hline
99 & Hyperfocal distance & Q253164 & $H = \frac{f^2}{Nc} + f$ & tp \\
\hline
100 & Coefficient of variation & Q623738 & $c_{v} = \frac{\sigma}{\mu}$ & tp \\
\hline
\end{tabular}
}
\egroup
\end{table}

\begin{table}
\caption{Contingency matrix of the general formula seeding.}
\label{tab:Contingency_general}
\centering
\begin{tabular}{|c|c|c|}
\hline
 & Relevant & Non-relevant\\
\hline
Retrieved & 71 (tp) & 17 (fp)\\
\hline
Not retrieved & 10 (fn) & 2 (tn)\\
\hline
\end{tabular}
\end{table}

We calculated the precision of the data transfer as

\begin{center}
P = 71/(71+17) = 0.8,
\end{center}

concluding that 80 \% of the retrieved results were relevant.

Furthermore, we calculated the recall as

\begin{center}
R = 71/(71+10) = 0.88,
\end{center}

concluding that 88 \% of the total relevant documents were successfully retrieved.

Finally, the combined (equally weighted) f-measure is

\begin{equation*}
F_1 = \frac{2 \cdot 0.8 \cdot 0.88}{0.8+0.88} = 0.84.
\end{equation*}

From this result, we can conclude that the seeding of general mathematical formulae from English Wikipedia articles to Wikidata yielded an overall accuracy of 84 \%.

\subsubsection{Geometry Questions.}

Eventually, we evaluated the accuracy of the formulae extracted from geometry related articles. The evaluation was carried out similar to the evaluation of general mathematics. However, due to the much smaller number of items, we did not choose a random sample but evaluated all the retrieved results, i.e., 65 formulae belonging to 49 Wikipedia articles.

Table \ref{tab:geometry_evaluation} contains some of the extracted formulae with their Wikipedia title and the corresponding contingency, whereas Table \ref{tab:Contingency_geometry} shows our evaulation results for the geometry formula seeding. The complete list is available at\\ \url{https://github.com/ag-gipp/MathQa}.

\begin{table}
\caption{Evaluation of the seeding of geometry formulae from Wikipedia to Wikidata.}
\label{tab:geometry_evaluation}
\centering
\bgroup
\def\arraystretch{1.8}
\resizebox{\columnwidth}{!}{
\begin{tabular}{|c|c|c|c|c|}
\hline
\textbf{Number} & \textbf{Wikipedia title} & \textbf{Property} & \textbf{Retrieved result} & \textbf{Contingency} \\
\hline
1 & Antiprism & Volume & $V = \frac{n \sqrt{4\cos^2\frac{\pi}{2n}-1}\sin \frac{3\pi}{2n} }{12\sin^2\frac{\pi}{n}}  a^3$ & tp \\
\hline
2 & Circle & Circumference & $C = 2 \pi r = \pi d$ & tp \\
 &  & Area & $\mathrm{Area} = \pi r^2$ & tp \\
\hline
 &  & Area & $A_\text{ellipse} = \pi ab$ & tp \\
... & ... & ... & ... & ... \\
\hline
48 & Law of cosines & Areas & $a^2 + b^2 = c^2 + 2ab \cos \gamma$ & fp \\
\hline
49 & Pentagon & Area & $A = \frac 1 2 Pr$ & tp \\
\hline
 & & & $(a^2 + b^2 - c^2)^2 \leq (4 A)^6$ & \\
 & & Circumradius & $R+r < \frac{a+b}{2}$ & fp \\
... & ... & ... & ... & ... \\
\hline
\end{tabular}
}
\egroup
\end{table}

\begin{table}
\caption{Contingency matrix of the geometry formula seeding.}
\label{tab:Contingency_geometry}
\centering
\begin{tabular}{|c|c|c|}
\hline
 & Relevant & Non-relevant\\
\hline
Retrieved & 52 (tp) & 1 (fp)\\
\hline
Not retrieved & 12 (fn) & 0 (tn)\\
\hline
\end{tabular}
\end{table}

Based on these values, we calculated the precision of the data transfer as\\

\begin{center}
P = 52/(52+1) = 0.98,
\end{center}

concluding that 98 \% of the retrieved results were relevant.

Furthermore, we calculated the recall as

\begin{center}
R = 52/(52+12) = 0.81,
\end{center}

concluding that 81 \% of the total relevant documents were successfully retrieved.

Finally, the combined (equally weighted) f-measure is

\begin{equation*}
F_1 = \frac{2 \cdot 0.98 \cdot 0.81}{0.98+0.81} = 0.87.
\end{equation*}

From this result, we can conclude that the seeding of geometry formulae from English Wikipedia articles to Wikidata yielded an overall accuracy of 87 \%.

\paragraph{Issues.}

Evaluating the seeding sample, we could observe some illustrative issue cases (see Table \ref{tab:SeedingIssues}) which will be briefly discussed in the following. There were some Wikipedia articles that did not contain a mathematical concept in the strong sense, but instead an algorithm (\# 4, 7, 48), measurement device (\# 35, 38), mathematical method or field (\# 25, 30, 33), a set (\# 49) or even a scientist (\# 9) or historical topic (\# 19). Some retrieved formulae (\# 6, 23) were only a part of the definition or statement which also contained natural language terms. We conclude that the \verb|<math>| tag is not a sufficient marker to find mathematical concepts within the bulk of Wikipedia articles. For future work, better filters will have to be developed that discard the articles mentioned above and possibly also other types we are currently not aware of.

\begin{table}
\resizebox{\columnwidth}{!}{
\begin{tabular}{|c|c|c|}
\hline
Evaluation \# and issue\\
\hline
4 \\ \textit{Blahut-Arimoto algorithm} is an \textit{algorithm}\\
\hline
6 \\ \textit{Quasi-finite field} formula is only a \textit{part of the definition}\\
\hline
7 \\ \textit{Streaming algorithm} is an \textit{algorithm}\\
\hline
9 \\ \textit{Carl Neumann} is a \textit{mathematician}\\
\hline
19 \\ \textit{Chinese mathematics} is a topic of the \textit{history of science}\\
\hline
20 \\ \textit{Bubble (physics)} pulsation formula \\ is only an \textit{example property} of a bubble\\
\hline
23 \\ \textit{Luce's choice axiom} formula is only a \textit{part of the statement}\\
\hline
25 \\ \textit{Sheaf cohomology} is the \textit{application} of homological algebra\\ to analyze the global sections of a sheaf on a topological space\\
\hline
30 \\ \textit{Enumerative combinatorics} is a field of mathematics\\
\hline
33 \\ \textit{Collocation method} is a method \\ (for the numerical solution of ordinary differential equations,\\ partial differential equations and integral equations)\\
\hline
35 \\ \textit{Lens clock} is a \textit{device} \\ (mechanical dial indicator that is used to measure \\ the dioptric power of a lens)\\
\hline
38 \\ \textit{Parshall flume} is (an open channel flow metering) \textit{device}\\
\hline
48 \\ \textit{Hierarchical network model} is an algorithm \\ (for creating networks which are able to reproduce \\ the unique properties \\ of the scale-free topology and the high clustering \\ of the nodes at the same time)\\
\hline
49 \\ \textit{Matching (graph theory)} is a set \\ (of edges without common vertices)\\
\hline
55 \\ renamed from \textit{Chemical milling} to \textit{Industrial etching}\\
\hline
\end{tabular}}
\caption{Issues we observed evaluating the seeding.}
\label{tab:SeedingIssues}
\end{table}

\subsection{Evaluation of the Math QA system}

Our math-aware QA system can answer mathematical questions in English and the Hindi language or use a direct formula input to deliver a calculation based on input values for the occurring identifiers.

\subsubsection{Evaluation of the Formula Retrieval Module.}

We evaluated our system on the basis of all formulae that were seeded correctly (true positive), determining whether a formula was retrieved (true or false) from Wikidata by the \textit{Formula Retrieval Module}.
The evaluation lists are available at\\ \url{https://github.com/ag-gipp/MathQa}. The retrieval of general mathematical formulae yielded 34 \textit{true} and 35 \textit{false} results. The accuracy of the system is

\begin{center}
Accuracy = Number of true results / Total size of the sample\\
= 34/(34+35) = 0.49
\end{center}

From the results, we can conclude that the ability to successfully retrieve a general mathematical formula possessed an accuracy of 49 \%. The retrieval of geometry formulae yielded an accuracy of 31 \%.
Our system can successfully answer questions provided by the user in the Hindi Language. So far, there is no tool available that can answer mathematical questions written in the Hindi language. So, we could not compare our results to any other tool.

\paragraph{Issues.}

Evaluating the MathQA sample, we could observe some illustrative issue cases (see Table \ref{tab:MathQAIssues}) which will be briefly stated in the following. Wikidata users renamed the \textit{has quality} property \textit{area} to \textit{area of plane shape}, which impeded our system from retrieving the respective formula. In some cases, there were too many synonymous Wikidata items available, so that the system could not filter out the requested mathematical concept. Furthermore, when processing the request \textit{Volume of a prism}, the system found \textit{prism - transparent optical element} (Q165896) instead of \textit{prism - geometric shape} (Q180544). Finally, if the name of the requested item contained an apostrophe ' or hyphen - it could not be processed properly.

\begin{table}
\begin{tabular}{|c|c|c|}
\hline
Evaluation \# and issue\\
\hline
1 \\ Wikidata item \\ has quality \textit{area of plane shape} \\ instead of just \textit{area}\\
\hline
3, 6, 9 \\ Wikidata has too many synonymous items available?\\
\hline
8 \\ system finds prism - transparent optical element (Q165896) \\ instead of prism - geometric shape (Q180544)\\
\hline
10 \\ Wikidata item missing\\
\hline
11, 13, 14, 17 \\ cause problems with ' ?\\
\hline
15, 32, 36 \\ link to disambiguation page\\
\hline
27 \\ enter value for \verb|\displaystyle|\\
\hline
48 \\ Wikidata has duplicate item\\
\hline
\end{tabular}
\caption{Issues we observed evaluating the QA system.}
\label{tab:MathQAIssues}
\end{table}

\subsubsection{Comparison to a commercial computational mathematical knowledge-engine}

Currently, there is no known tool available which delivers a direct formula answer and performs a calculation using input values for the identifiers. So we could not fully compare our results to other systems. However, we studied the ability of our system to successfully answer questions compared to a selected computational knowledge engine that performs arithmetic operations and answers questions from different fields of general knowledge.

\begin{table}
\centering
\bgroup
\def\arraystretch{1.8}
\resizebox{\columnwidth}{!}{
\begin{tabular}{|l|}
\hline
\begin{tabular}{p{0.6\columnwidth}p{0.5\columnwidth}p{0.15\columnwidth}}
Our system & Commercial engine & Perf.
\end{tabular}
\\
\hline
\textbf{1. What is the formula for Riemann zeta function?}\\
\begin{tabular}{p{0.6\columnwidth}p{0.5\columnwidth}p{0.05\columnwidth}}
$\zeta(s) \sum_{n=1}^\infty \frac{1}{n^s}$ & $\zeta(s) = 2^s \pi^{(-1 + s)} \Gamma(1 - s) \sin((\pi s)/2) \zeta(1 - s)$ & \ \ \ =
\end{tabular}
\\
\hline
\textbf{2. What is the formula for Logical equivalence?}\\
\begin{tabular}{p{0.6\columnwidth}p{0.5\columnwidth}p{0.05\columnwidth}}
$p \iff q$ & No result & \ \ \ $>$
\end{tabular}
\\
\hline
\textbf{3. What is the formula for Direct sum?}\\
\begin{tabular}{p{0.6\columnwidth}p{0.5\columnwidth}p{0.05\columnwidth}}
$(a_1, b_1) \cdot (a_2, b_2) = (a_1 \circ a_2, b_1 \bullet b_2)$ &
$\sum_{k=0}^\infty x^k = 1/(1 - x) \ \mathrm{when} \ abs(x)<1$
& \ \ \ $>$
\end{tabular}
\\
\hline
\textbf{4. What is the formula for Determinant?}\\
\begin{tabular}{p{0.6\columnwidth}p{0.5\columnwidth}p{0.05\columnwidth}}
$\det(A) \sum_{\sigma \in S_n} ( (\sigma) \prod_{i=1}^n a_{i,\sigma_i})$ & Computation of 3x3 determinant & \ \ \ =
\end{tabular}
\\
\hline
\textbf{5. What is the formula for Hypergeometric function?}\\
\begin{tabular}{p{0.6\columnwidth}p{0.5\columnwidth}p{0.05\columnwidth}}
${}_2F_1(a,b;c;z) \sum_{k=0}^\infty \frac{ \Gamma(a+k) \, \Gamma(b+k) \, \Gamma(c) }{ \Gamma(a) \,\Gamma(b)\, \Gamma(c+k)} \frac{z^k}{k!}$ & >
$c_{k+1}/c_k = P(k)/Q(k) = ...$
& \ \ \ =
\end{tabular}
\\
\hline
\textbf{6. What is the formula for Continued fraction?}\\
\begin{tabular}{p{0.6\columnwidth}p{0.5\columnwidth}p{0.05\columnwidth}}
$a_0 + \cfrac{b_1}{a_1 + \cfrac{b_2}{a_2 + \cfrac{b_3}{a_3 + ...}}}$ & 3.17 & \ \ \ $>$
\end{tabular}
\\
\hline
...
\\
\hline
\textbf{26. What is the formula for Pythagorean theorem?}\\
\begin{tabular}{p{0.6\columnwidth}p{0.5\columnwidth}p{0.05\columnwidth}}
$c^2 = a^2 + b^2$ & $a^2 + b^2 = c^2$ & \ \ \ =
\end{tabular}
\\
\hline
\textbf{27. What is the formula for Pythagorean triple?}\\
\begin{tabular}{p{0.6\columnwidth}p{0.5\columnwidth}p{0.05\columnwidth}}
$a = m^2 - n^2, b = 2mn, c = m^2 + n^2$ & $a^2 + b^2 = c^2$ & \ \ \ $>$
\end{tabular}
\\
\hline
\textbf{28. What is the formula for Jensen's inequality?}\\
\begin{tabular}{p{0.6\columnwidth}p{0.5\columnwidth}p{0.05\columnwidth}}
No result & $f(\sum_{i = 1}^n p_i x_i) <= \sum_{i = 1}^n p_i f(x_i)$ if $f$ convex\\
 & $f(\sum_{i = 1}^n p_i x_i) >= \sum_{i = 1}^n p_i f(x_i)$ if $f$ concave
 & \ \ \ $<$
\end{tabular}
\\
\hline
\textbf{29. What is the formula for Logistic function?}\\
\begin{tabular}{p{0.6\columnwidth}p{0.5\columnwidth}p{0.05\columnwidth}}
$f(x) = \frac{L}{1 + \mathrm e^{-k(x-x_0)}}$ & $1/(1 + exp(-x)) = e^x/(1 + e^x)$ & \ \ \ $>$
\end{tabular}
\\
\hline
\textbf{30. What is the formula for Convolution?}\\
\begin{tabular}{p{0.6\columnwidth}p{0.5\columnwidth}p{0.05\columnwidth}}
$(f*g)(t) \stackrel{\mathrm{def}}{=} \int_{-\infty}^\infty f(\tau) g(t - \tau) d\tau$ \\ $= \int_{-\infty}^\infty f(t-\tau) g(\tau) d\tau$ & No result & \ \ \ $>$
\end{tabular}
\\
\hline
\textbf{Evaluation results:}\\
14 = , 10 $>$, 6 $<$
\\
\hline
\end{tabular}
}
\egroup
\vspace{0.5cm}
\caption{NTCIR 12 \cite{DBLP:conf/ntcir/SchubotzMLG16} example questions in which our system is able to outperform the selected commercial engine.}
\label{tab:PerformanceComparison}
\end{table}
Quantitatively, we used 30 mathematical questions from the NTCIR-12 Task \cite{DBLP:conf/ntcir/SchubotzMLG16} for an evaluation of the performance of the two systems. After approving or seeding 5 missing formulae manually, our system was able to outperform the commercial engine, yielding more suitable answers (denoted by $>$ in Table \ref{tab:PerformanceComparison} column Performance (Perf.)) in 10/30 of the cases. The reference engine performed better than our system in only 6/30 (denoted by $<$), and in 14/30 questions both systems provided answers that were estimated to be equally suitable (denoted by $=$). All in all, our system was able to outperform the commercial engine on the NTCIR-12 sample. Nevertheless, it should be mentioned that our reference engine is continuously striving to improve on mathematical topics. For example, the question \textit{What is the formula for Logical equivalence?} yielded \textit{Additional functionality for this topic is under development...} and we suspect that there are more of these cases.
Qualitatively, we could observe that our system is more powerful in comparison to the reference engine when answering definition questions. As an example, the question \textit{What is the formula for gas?} is answered by $PV=nRT$, whereas the reference engine only returns a list of gaseous compounds. Furthermore, our system can successfully answer geometry questions, whereas the reference engine provides all formulae with unit edge length and is not giving any option to enter a customized edge length. For example, the question \textit{What is the surface area of triangular cupola?} is answered as $A = \left(\frac{5 \sqrt 3}{2}\right) a^2$, whereas the reference engine only displays $3+ \frac{5 \sqrt 3}{2} \approx 7.33013$. However, our QA system can answer only mathematical questions, whereas the reference engine can answer questions from many other fields like people and history, health and medicine, materials, dates and times, engineering, earth science, etc.

\paragraph{Limitations.}

Our system decisively depends on the knowledge stored in Wikidata. If the item or the formula we are looking for is not available in Wikidata, we are unable to answer a given question. Furthermore, we are using the question parsing module developed by Platypus [10] which is limited to the use of nouns in singular form leading to an inability to answer a question containing a plural noun. For instance, the question \textit{What is the formula for Maxwell's equations?} is parsed as \textit{Maxwell's equation} such that the item cannot be retrieved. Besides, our parser does not support specific LaTeX tags (\verb|\displaystyle,| \verb|\frac,| \verb|\left,| \verb|\right,| \verb|\bigg,| \verb|\mathrm,| etc.) or punctuation symbols (, ; ! etc.) as well as integration, summation, scalar products or more complex formulae. For the Hindi language questions, we are limited to the available Wikidata items that include Hindi labels. So, we are unable to process all the Wikidata items available in English in the Hindi language also.

\section{Conclusion}

The overall goal of this research project was to extract mathematical knowledge in the form of formulae from English Wikipedia and seed it into Wikidata. This served as a necessary preparation for building a QA system that can answer mathematical questions in English and Hindi language. Additionally, the user can perform arithmetic calculations using the retrieved formula, after providing input values for the detected identifiers. We have been able to provide the Wikidata community with more than 17 thousand new Wikidata statements containing formulae. Our seeding achieved a precision of 80 \%, recall of 88 \% and a combined f-measure of 84 \% for general mathematical formulae. For the geometry formulae, the precision was 98 \%, the recall 81 \% and the f-measure yielded 87 \%. The \textit{Formula Retrieval Module} of our QA system possessed an accuracy of 49 \% and 31 \% for general and geometry formulae respectively. As far as we are concerned, our QA system is the only available to answer mathematical questions in the Hindi language.

\paragraph{Future Work.}

Wikipedia is the world's largest online encyclopedia and consists of a massive amount of information in numerous different languages. There are many possibilities when it comes to the task of migrating data from Wikipedia to Wikidata. This research project was only dealing with mathematical knowledge in Wikipedia. However, similar techniques can also be employed to migrate knowledge from other fields such as geography, computer science, politics and many more between these Wiki sister projects.
We propose the following possibilities for future work: 1.) Use of another database or knowledge-base than Wikidata or adding a module that can use another database if the requested item is not available in Wikidata. 2.) Seeding Wikidata with more mathematical formulae to enable answering more questions. 3.) Seeding Wikidata with more Hindi labels for Wikidata items to improve the performance of our Hindi language module. 4.) Develop a new LaTeX parser that can parse any latex formula without restriction. 5.) Improve the Formula Retrieval Module allowing for plots and more information regarding the formula. 6.) Improve the Formula Calculation Module such that it delivers the calculated result including units of the formula and the identifiers.

\begin{acks}
We would like to thank Akiko Aizawa for her advice and for hosting us as visiting researchers in her lab at the National Institute of Informatics (NII) in Tokyo. Furthermore, we thank Wikimedia Foundation and Wikimedia Deutschland providing cloud computing facilities and us for a research visit. Besides many Wikimedians, Lydia Pintscher and Jonas Kress were a great help in getting started with Wikidata.
This work was supported by the FITWeltweit program of the German Academic Exchange Service (DAAD) as well as the \grantsponsor{DFG}{German Research Foundation (DFG}, grant \grantnum{DFG}{GI-1259-1}).
\end{acks}

\printbibliography[keyword=primary]
\end{document}